\documentclass[12pt]{article}
\pdfoutput=1
 \def\be{\begin{equation}}
 \def\ee{\end{equation}}
 \def\bea{\begin{eqnarray}}
 \def\eea{\end{eqnarray}}
 \usepackage{graphicx}

 \catcode`\@=11
 \def\lsim{\mathrel{\mathpalette\@versim<}}
 \def\gsim{\mathrel{\mathpalette\@versim>}}
 \def\@versim#1#2{\vcenter{\offinterlineskip
 \ialign{$\m@th#1\hfil##\hfil$\crcr#2\crcr\sim\crcr } }}
 \catcode`\@=12

 \catcode`@=12
\begin{document}
 \thispagestyle{empty}
 \begin{flushright}
 UCRHEP-T566\\
 May 2016\
 \end{flushright}
 \vspace{0.6in}
 \begin{center}
 {\LARGE \bf Gauge $B-L$ Model of Radiative Neutrino\\
 Mass with Multipartite Dark Matter\\}
 \vspace{1.2in}
 {\bf Ernest Ma, Nicholas Pollard, Oleg 
Popov, and Mohammadreza Zakeri\\}
 \vspace{0.2in}
 {\sl Department of Physics and Astronomy,\\ 
 University of California, Riverside, California 92521, USA\\}
 \end{center}
 \vspace{1.2in}

\begin{abstract}\
We propose an extension of the standard model of quarks and leptons to 
include gauge $B-L$ symmetry with an exotic array of neutral fermion 
singlets for anomaly cancellation.  With the addition of suitable 
scalars also transforming under $U(1)_{B-L}$, this becomes a model 
of radiative seesaw neutrino mass with possible multipartite dark matter.  
If leptoquark 
fermions are added, necessarily also transforming under $U(1)_{B-L}$, 
the diphoton excess at 750 GeV, recently observed at the Large Hadron 
Collider, may also be explained.    
\end{abstract}

 \newpage
 \baselineskip 24pt

\noindent \underline{\it Introduction}~:\\
It is well-known that a gauge $B-L$ symmetry is supported by a simple 
extension of the standard model (SM) of quarks and leptons with the addition 
of one singlet right-handed neutrino per family, so that the theory is 
anomaly-free.  For convenience in notation, let these three extra neutral 
fermion singlets $N$ be left-handed, then their charges under $U(1)_{B-L}$ 
are (1,1,1).  Their additional contributions to the axial-vector anomaly 
and the mixed gauge-gravitational anomaly are respectively 
\begin{equation}
(1)^3 + (1)^3 + (1)^3 = 3, ~~~ (1) + (1) + (1) = 3,
\end{equation}
which cancel exactly those of the SM quarks and leptons.  On the other hand, 
it has been known for some time~\cite{mp09} that another set of charges 
are possible, i.e.
\begin{equation}
(-5)^3 + (4)^3 + (4)^3 = 3, ~~~ (-5) + (4) + (4) = 3.
\end{equation}
Adding also three pairs of neutral singlet fermions with charges $(1,-1)$, 
naturally small seesaw Dirac masses for the known three neutrinos may be 
obtained~\cite{ms15}, and a residual global $U(1)$ symmetry is maintained 
as lepton number.  A further extension in the scalar sector allows for 
the unusual case of $Z_3$ lepton number~\cite{mpsz15} with the appearance of 
a scalar dark-matter candidate which is unstable but long-lived and decays 
to two antineutrinos.  Here we consider another set of possible charges 
for the neutral fermion singlets, such that tree-level neutrino masses are 
forbidden.  New scalar particles transforming under $U(1)_{B-L}$ are then 
added to generate one-loop Majorana neutrino masses.  The breaking of 
$B-L$ to $Z_2$ results in lepton parity and thus $R$ parity or dark 
parity~\cite{m15} which is odd for some particles, the lightest neutral 
one being dark matter.  A closer look st the neutral fermion singlets shows 
that one may be a keV sterile neutrino, and two others are heavy and 
stable, thus realizing the interesting scenario of multipartite dark matter.  
If color-triplet fermions with both $B$ and $L$ 
are added, the diphoton excess~\cite{atlas15,cms15} at 750 GeV, recently 
observed at the Large Hadron Collider (LHC), may also be explained.

\noindent \underline{\it Model}~:\\
The extra left-handed neutral singlet fermions have charges 
$(2,2,2,2,-1,-1,-3)$, so that
\begin{equation}
4(2)^3 + 2(-1)^3 + (-3)^3 = 3, ~~~ 4(2) + 2(-1) + (-3) = 3.
\end{equation}
Since there is no charge +1 in the above, there is no connection between them 
and the doublet neutrinos $\nu$ with charge $-1$ through the one Higgs doublet 
$\Phi$ which has charge zero.  Neutrinos are thus massless at tree level.  To 
generate one-loop Majorana masses, the basic mechanism of Ref.~\cite{m06} 
is adopted, using the four fermions with charge +2, but because of the 
$U(1)_{B-L}$ gauge symmetry, we need both a scalar doublet $(\eta^+,\eta^0)$ 
and a scalar singlet $\chi^0$.
\begin{table}[htb]
\caption{Particle content of proposed model.}
\begin{center}
\begin{tabular}{|c|c|c|c|c|c|c|c|c|}
\hline
Particle & $SU(3)_C$ & $SU(2)_L$ & $U(1)_Y$ & $B$ & $L$ & $B-L$ & copies & 
$R$ parity\\
\hline
$Q = (u,d)$ & 3 & 2 & 1/6 & 1/3 & 0 & 1/3 & 3 & + \\
$u^c$ & $3^*$ & 1 & $-2/3$ & $-1/3$ & 0 & $-1/3$ & 3 & + \\
$d^c$ & $3^*$ & 1 & 1/3 & $-1/3$ & 0 & $-1/3$ & 3 & + \\
\hline
$L = (\nu,e)$ & 1 & 2 & $-1/2$ & 0 & 1 & $-1$ & 3 & + \\
$e^c$ & 1 & 1 & 1 & 0 & $-1$ & 1 & 3 & + \\
\hline
$N$ & 1 & 1 & 0 & 0 & $-2$ & 2 & 4 & $-$ \\  
$S$ & 1 & 1 & 0 & 0 & 1 & $-1$ & 2 & + \\
$S'$ & 1 & 1 & 0 & 0 & 3 & $-3$ & 1 & + \\ 
\hline
$\Phi = (\phi^+,\phi^0)$ & 1 & 2 & $1/2$ & 0 & 0 & 0 & 1 & + \\
$\eta = (\eta^+,\eta^0)$ & 1 & 2 & 1/2 & 0 & 1 & $-1$ & 1 & $-$ \\
$\chi^0$ & 1 & 1 & 0 & 0 & 1 & $-1$ & 1 & $-$ \\
$\rho^0_2$ & 1 & 1 & 0 & 0 & 2 & $-2$ & 1 & + \\
$\rho^0_4$ & 1 & 1 & 0 & 0 & 4 & $-4$ & 1 & + \\
\hline
$D_1$ & 3 & 1 & $-1/3$ & 1/3 & 1 & $-2/3$ & 1 & $-$ \\
$D_2$ & 3 & 1 & $-1/3$ & 1/3 & $-1$ & 4/3 & 1 & $-$ \\
$D^c_1$ & $3^*$ & 1 & 1/3 & $-1/3$ & $-1$ & 2/3 & 1 & $-$ \\
$D^c_2$ & $3^*$ & 1 & 1/3 & $-1/3$ & 1 & $-4/3$ & 1 & $-$ \\
\hline
\end{tabular}
\end{center}
\end{table}
The $U(1)_{B-L}$ gauge symmetry itself is broken by $\rho_2^0$ with charge 
$-2$ and by $\rho_4^0$ with charge $-4$.  The leptoquark fermions $D_{1,2}$ 
and $D^c_{1,2}$ are not necessary for neutrino mass, but are natural 
extensions of this model if the diphoton excess at 750 GeV requires an 
explanation. The complete particle content of this model is shown in Table 1.

\noindent \underline{\it Radiative Neutrino Mass}~:\\
Using the four $N$'s, radiative Majorana masses for the three $\nu$'s are 
generated as shown in Fig.~1.
\begin{figure}[htb]
\vspace*{-2cm}
\hspace*{-3cm}
\includegraphics[scale=1.0]{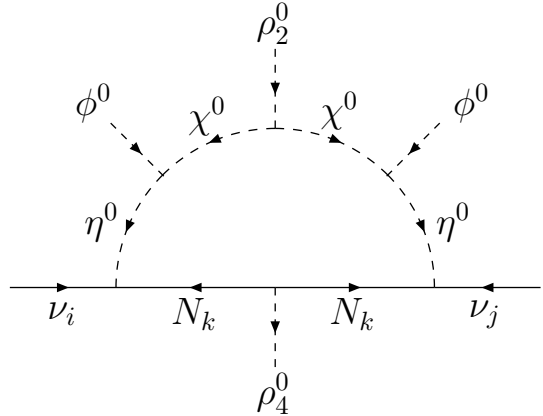}
\vspace*{-21cm}
\caption{Radiative generation of neutrino mass through dark matter.}
\end{figure}
Note that $N,\eta,\chi$ all have odd $R$ parity, so that the lightest neutral 
particle among them is a dark-matter candidate.  This is the scotogenic 
mechanism, from the Greek 'scotos' meaning darkness.  In addition to the 
$\eta^\dagger \Phi \chi$ trilinear coupling used in Fig.~1, there is also 
the $\eta^\dagger \Phi \chi^\dagger \rho_2$ quadrilinear coupling, which 
may also be used to complete the loop.  There are 4 real scalar fields, 
spanning $\sqrt{2}Re(\eta^0)$, $\sqrt{2}Im(\eta^0)$, $\sqrt{2}Re(\chi^0)$, 
$\sqrt{2}Im(\chi^0)$.  We denote their mass eigenstates as $\zeta^0_l$ 
with mass $m_l$.  Let the $\nu_i N_k \eta^0$ coupling be $h^\nu_{ik}$, then 
the radiative neutrino mass matrix is given by~\cite{m06}
\begin{equation}
({\cal M}_\nu)_{ij} = \sum_k {h^\nu_{ik} h^\nu_{jk} M_k \over 16 \pi^2} 
\sum_l [(y^R_l)^2 F(x_{lk}) - (y^I_l)^2 F(x_{lk})],
\end{equation}
where $\sqrt{2}Re(\eta^0) = \sum_l y_l^R \zeta_l^0$, 
$\sqrt{2}Im(\eta^0) = \sum_l y_l^I \zeta_l^0$, with $\sum_l (y_l^R)^2 = 
\sum_l (y_l^I)^2 = 1$, $x_{lk} = m_l^2/M_k^2$, and the function $F$ is given by
\begin{equation}
F(x) = {x \ln x \over x-1}.
\end{equation}

\noindent \underline{\it Multipartite Dark Matter}~:\\
Since the only neutral particles of odd $R$ parity are $N,\eta^0,\chi^0$, 
there appears to be only one dark-matter candidate.  However as shown 
below, there could be two or even four, all within the context of the 
existing model.

First note that $\rho^0_{2,4}$ have exactly the right $U(1)_{B-L}$ charges to 
make the $(S,S,S')$ fermions massive.  The corresponding $3 \times 3$ mass 
matrix is of the form
\begin{equation}
{\cal M}_S = \pmatrix{m_{S1} & 0 & m_{13} \cr 0 & m_{S2} & m_{23} \cr 
m_{13} & m_{23} & 0},
\end{equation}
where $m_{S1},m_{S2}$ come from $\langle \rho^0_2 \rangle = u_2$ and 
$m_{13},m_{23}$ from $\langle \rho^0_4 \rangle = u_4$.  If all these 
entries are of order 100 GeV to a few TeV, then there are three extra 
heavy singlet neutrinos in this model which also have even $R$ parity. 
They do not mix with the light active neutrinos $\nu$ at tree level, 
but do so in one loop.  For example, $S'$ mixes with $\nu$ as shown 
in Fig.~2.
\begin{figure}[htb]
\vspace*{-3cm}
\hspace*{-3cm}
\includegraphics[scale=1.0]{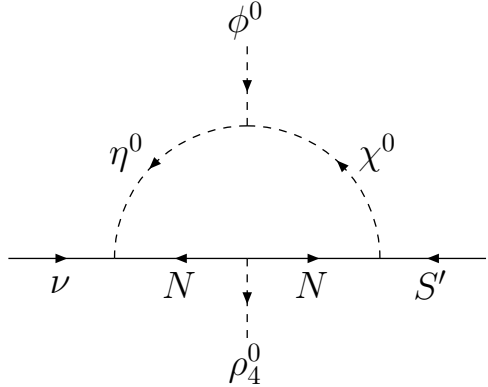}
\vspace*{-20.5cm}
\caption{Radiative generation of $\nu-S'$ mixing.}
\end{figure}
Similarly $S$ will also mix with $\nu$, using the $S N \chi^0$ Yukawa 
coupling.  However, these terms are negligible compared to the assumed 
large masses for $(S,S,S')$ and may be safely ignored.

Consider now the possibility that $m_{13},m_{23} << m_{S1},
m_{S2}$ in ${\cal M}_S$, then $S'$ obtains a small seesaw mass given by
\begin{equation}
m_{S'} \simeq - {m_{13}^2 \over m_{S1}} - {m_{23}^2 \over m_{S2}}.
\end{equation}
Let this be a few keV, then $S'$ is a light sterile neutrino which mixes 
with $\nu$ only slightly through Fig.~2.  Hence it is a candidate for 
warm dark matter.  Whereas the usual sterile neutrino is an {\it ad hoc} 
invention, it has a natural place here in terms of its mass as well as 
its suppressed mixing with the active neutrinos.

We now have the interesting scenario where part of the dark matter of the 
Universe is cold, and the other is warm.  This hybrid case was recently 
also obtained in a different radiative model of neutrino masses~\cite{abm16}. 
Within the present context, there is a third possibility.  If we assign an 
extra $Z_2$ symmetry, under which $S_{1,2}$ are odd and all other particles 
even, then the only interactions involving $S_{1,2}$ come from their diagonal 
$U(1)_{B-L}$ gauge couplings and the diagonal Yukawa terms 
$f_1 S_1 S_1 (\rho^0_2)^*$ and $f_2 S_2 S_2 (\rho^0_2)^*$.  This means 
that both $S_1$ and $S_2$ are stable and their relic abundances are 
determined by their annihilation cross sections to SM particles.  
In this scenario, dark matter has four components~\cite{cmwy07}.

Since $S_{1,2}$ are now separated from $S'$, the $m_{13}$ and $m_{23}$ terms 
in ${\cal M}_S$ are zero and there is no tree-level mass for $S'$. 
However, there is a one-loop mass as shown in Fig.~3.
\begin{figure}[htb]
\vspace*{-3cm}
\hspace*{-3cm}
\includegraphics[scale=1.0]{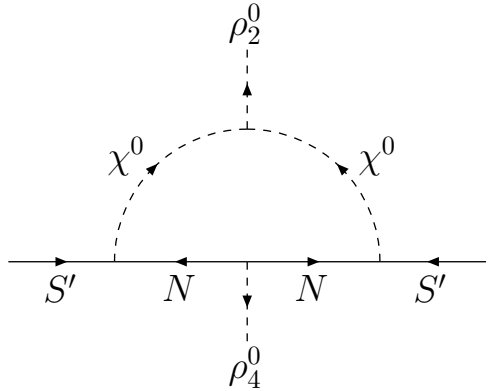}
\vspace*{-20.5cm}
\caption{Radiative generation of $S'$ mass.}
\end{figure}
This makes it more natural for $S'$ to be light.  A detailed study of the 
dark-matter phenomenology of this multipartite scenario will be given 
elsewhere.

\newpage
\noindent \underline{\it Scalar Sector for Symmetry Breaking}~:\\
In this model, there is only one Higgs doublet $\Phi$ which breaks the 
$SU(2)_L \times U(1)_Y$ electroweak symmetry, whereas there are two 
Higgs singlets $\rho_2$ and $\rho_4$ which break $U(1)_{B-L}$ to $Z_2$. 
The most general Higgs potential consisting of $\Phi,\rho_2,\rho_4$ is 
given by
\begin{eqnarray}
V &=& \mu_0^2 \Phi^\dagger \Phi + \mu_2^2 \rho_2^* \rho_2 + \mu_4^2 
\rho_4^* \rho_4 + {1 \over 2} \mu_{24} [\rho_2^2 \rho_4^* + H.c.] 
+ {1 \over 2} \lambda_0 (\Phi^\dagger \Phi)^2 + {1 \over 2} \lambda_2 
(\rho_2^* \rho_2)^2 \nonumber \\ 
&+& {1 \over 2} \lambda_4 (\rho_4^* \rho_4)^2 +  
\lambda_{02} (\Phi^\dagger \Phi)(\rho_2^* \rho_2) + \lambda_{04} 
(\Phi^\dagger \Phi)(\rho_4^* \rho_4) + \lambda_{24} (\rho_2^* \rho_2) 
(\rho_4^* \rho_4).
\end{eqnarray}
Let $\langle \phi^0 \rangle = v$, $\langle \rho_2 \rangle = u_2$, 
$\langle \rho_4 \rangle = u_4$, then the minimum of $V$ is determined by
\begin{eqnarray}
0 &=& \mu_0^2 + \lambda_0 v^2 + \lambda_{02} u_2^2 + \lambda_{04} u_4^2, \\ 
0 &=& \mu_2^2 + \lambda_{02} v^2 + \lambda_2 u_2^2 + \lambda_{24} u_4^2 
+ \mu_{24} u_4, \\ 
0 &=& u_4(\mu_4^2 + \lambda_{04} v^2 + \lambda_{24} u_2^2 + \lambda_4 u_4^2) 
+ {1 \over 2} \mu_{24} u_2^2.
\end{eqnarray}
The would-be Goldstone bosons are $\phi^\pm$, $\sqrt{2}Im(\phi^0)$, 
corresponding to the breaking of $SU(2)_L \times U(1)_Y$ to $U(1)_{em}$, 
and $\sqrt{2}[u_2 Im(\rho_2) + 2u_4 Im(\rho_4)]/\sqrt{u_2^2 + 4u_4^2}$, 
corresponding to the breaking of $U(1)_{B-L}$ to $Z_2$.  The 
linear combination orthogonal to the latter is a physical pseudoscalar $A$, 
with a mass given by
\begin{equation}
m_A = {-\mu_{24} (u_2^2 + 4 u_4^2) \over 2 u_4}.
\end{equation}
The $3 \times 3$ mass-squared matrix of the physical scalars 
$[\sqrt{2}Re(\phi^0), \sqrt{2}Re(\rho_2), \sqrt{2}Re(\rho_4)]$ 
is given by
\begin{equation}
{\cal M}^2 = \pmatrix{2 \lambda_0 v^2 & 2 \lambda_{02} v u_2 & 
2 \lambda_{04} v u_4 \cr 2 \lambda_{02} v u_2 & 2 \lambda_2 u_2^2 & 
u_2(2\lambda_{24} u_4 + \mu_{24}) \cr 2 \lambda_{04} v u_4 & 
u_2 (2 \lambda_{24} u_4 + \mu_{24}) & 2 \lambda_4 u_4^2 - \mu_{24} u_2^2/2u_4}.
\end{equation}
For $v^2 << u^2_{2,4}$, $\sqrt{2}Re(\phi^0)=h$ is approximately a mass 
eigenstate which is identified with the 125 GeV particle discovered 
at the LHC.

\newpage
\noindent \underline{\it Gauge Sector}~:\\
Since $\phi^0$ does not transform under $U(1)_{B-L}$ and $\rho_{2,4}$ do 
not transform under $SU(2)_L \times U(1)_Y$, there is no tree-level mixing 
between their corresponding gauge bosons $Z$ and $Z_{B-L}$.  In our convention, 
$M^2_{Z_{B-L}} = 8 g_{B-L}^2 (u_2^2 + 4u_4^2)$.  The LHC bound on $M_{Z_{B-L}}$ comes 
from the production of $Z_{B-L}$ from $u$ and $d$ quarks and its subsequent 
decay to $e^- e^+$ and $\mu^- \mu^+$.  If all the particles listed in Table 1 
are possible decay products of $Z_{B-L}$ with negligible kinematic suppression, 
then its branching fraction to $e^- e^+$ and $\mu^- \mu^+$ is about 0.061. 
The $c_{u,d}$ coefficients used in the LHC analysis~\cite{atlas14,cms14} are then
\begin{equation}
c_u = c_d = \left[ \left( {1 \over 3} \right)^2 + \left( {1 \over 3} \right)^2 
\right] g^2_{B-L} \times B(Z_{B-L} \to e^-e^+, \mu^-\mu^+) = 1.36 \times 
10^{-2} ~g^2_{B-L}.
\end{equation}
From LHC data based on the 7 and 8 TeV runs, a bound of about 2.5 TeV would 
correspond to $g_{B-L} < 0.24$.

\noindent \underline{\it Leptoquark Fermions}~:\\
The singlet leptoquark fermions $D_{1,2}$ have charge $-1/3$ and the 
following possible interactions:
\begin{equation}
D_1 d^c \chi^*, ~~~ D_2 d^c \chi, ~~~ D_1 D_2^c \rho_2^*, ~~~ D_2 D_1^c \rho_2.
\end{equation}
Hence they mix in a $2 \times 2$ mass matrix linking $D_{1,2}$ to $D^c_{1,2}$ 
with $\langle \rho_2 \rangle = u_2$, and decay to $d$ quarks + $\chi(\chi^*)$. 
Now $\chi$ mixes with $\eta^0$, so it decays to neutrinos ($\nu$) and dark 
matter ($N$), which are invisible.  The search for $D_{1,2}$ at the LHC 
would be similar to the search for scalar quarks which decay to quarks + 
missing energy.  However, if we assume that $N$ has a mass of about 200 
GeV, then there is no useful limit at present on the mass of $D_{1,2}$ 
from the LHC. 

Consider now the pseudoscalar $A$ of Eq.~(12).  Let the two mass eigenstates 
in the $(D_{1,2},D^c_{1,2})$ sector be $\psi_{1,2}$, then $A$ couples to them 
according to
\begin{equation}
{\cal L}_{int} = f_1 \bar{\psi}_1 \gamma_5 \psi_1 + f_2 \bar{\psi}_2 
\gamma_5 \psi_2, 
\end{equation}
where $f_{1,2}$ are rearranged from their original $D_1 D_2^c \rho_2^*$ and 
$D_2 D_1^c \rho_2$ couplings.  Hence $A$ decays to two gluons as well 
as to two photons in one loop through $\psi_{1,2}$.  It may also decay to 
dark matter, say $NN$, at tree level.  It is thus a possible candidate 
for explaining the 750 GeV diphoton excess recently 
observed~\cite{atlas15,cms15} at the LHC.  The numerical analysis 
of this model runs parallel to that of a recent proposal~\cite{fkmppz16}, 
and will not be repeated here.  Note again that these leptoquark fermions 
are not essential for the radiative generation of neutrino masses based 
on $B-L$. 

\noindent \underline{\it Conclusion}~:\\
Using gauge $U(1)_{B-L}$ symmetry, we have proposed a new anomaly-free solution 
with exotic fermion singlets, such that neutrino mass is forbidden at tree 
level.  We add a number of new scalars so that neutrino masses are obtained 
in one loop through dark matter, i.e. the scotogenic mechanism.  Because 
of the structure of the new singlets required for anomaly cancellation, we 
find a possible dark-matter scenario with four components.  Three are 
stable cold Weakly Interaction Massive Particles (WIMPs) and one a keV 
singlet neutrino, i.e. warm dark matter with a very long lifetime. 
If leptoquark fermions are added, transforming under $U(1)_{B-L}$, the 
recently observed 750 GeV diphoton excess may also be explained.

\bigskip

\noindent \underline{\it Acknowledgement}~:~
This work was supported in part by the U.~S.~Department of Energy Grant 
No. DE-SC0008541.

\bibliographystyle{unsrt}

\end{document}